\documentclass[prb,aps,showpacs,twocolumn,unsortedaddress]{revtex4}
\usepackage{graphics,bm}
\usepackage{amssymb}
\usepackage{epsfig}
\usepackage{epsf}
\begin{document}

\title{Edge Magnetoplasmons in Quantum Hall Line Junction Systems}

\author{Wei-Cheng Lee}
\email{leewc@mail.utexas.edu}

\author{N. A. Sinitsyn}

\author{Emiliano Papa}

\author{A.H. MacDonald}
\email{macd@physics.utexas.edu}
\homepage{http://www.ph.utexas.edu/~macdgrp}

\affiliation{Department of Physics, The University of Texas at Austin, Austin, TX 78712}

\date{\today}

\begin{abstract}
A quantum Hall line junction system consists of a one-dimensional Luttinger 
liquid (LL) and two chiral channels that allow density waves incident upon and 
reflected by the LL to be measured separately.  We demonstrate that interactions in a quantum Hall
line junction system can be probed by studying edge magnetoplasmon absorption spectra
and their polarization dependences.  Strong interactions in
the junction lead to collective modes that are isolated in either Luttinger liquid or 
contact subsystems.  
\end{abstract}

\maketitle

\def\bx{{\bf x}}
\def\bk{{\bf k}}
\def\half{\frac{1}{2}}
\def\args{(\bx,t)}

\begin{figure}
\includegraphics{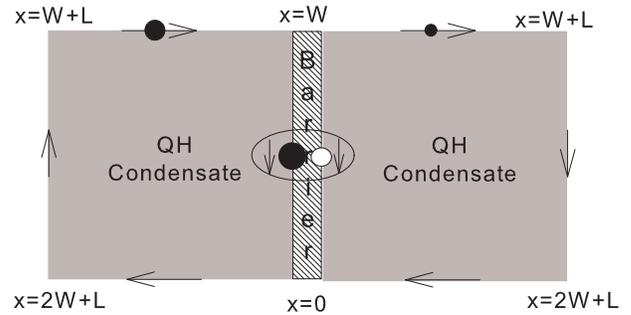}
\caption{\label{fig:one} Cartoon of a QH line junction.  The circles and arrows
illustrate, for the case of repulsive interactions of moderate strength,
the evolution of a unit charge incident on the top of the LL discussed in the text.   
The total perimeter of each QH region is $P=2W+2L$ and, in the coordinate system we 
use, the barrier region is the interval $[0,W]$.}
\end{figure}

\noindent {\it Introduction} --- A quantum Hall line junction (QHLJ)\cite{RennLJ,KaneLJ,ZulickeLJ}
enables the realization of one-dimensional electron systems with widely tunable properties.
A line junction is generated by creating a narrow 
barrier that divides a two-dimensional electron system on a quantum Hall plateau into separate subsystems
as illustrated in Fig.~\ref{fig:one}. 
Chiral quantum Hall edge channels \cite{halperinedge,wenedge,macdedge} flow in opposite directions 
on opposite sides of the barrier and constitute a non-chiral one-dimensional electron gas.
An attractive feature of QHLJ systems is the physical separation of incident and reflected
states at the ends of the non-chiral barrier region, a benefit provided by the chiral 
quantum Hall edge states.  This feature plays a role, for example, in proposed electron 
teleportation effects\cite{beenakker} on quantum Hall edges.

The narrow barriers that define QHLJs can be realized by cleaved edge overgrowth\cite{KangLJ},
corner overgrowth\cite{GraysonLJ}, or by the deposition of narrow metallic gates\cite{PisaLJ,HeiblumLJ,SaclayLJ}.
Experimental studies of various QHLJ systems, and some of the 
theoretical analyses\cite{MitraLJ,SachdevLJ,FradkinLJ,PapaLJ} that they have motivated, 
have made it clear that interactions between electrons on opposite sides of the 
barrier can play an essential role in their physics.  An 
important difficulty that arises in interpreting the transport properties of QHLJ systems 
is uncertainty about the strength and sometimes even the sign of these interactions,
which can be difficult to estimate because 
of subtleties\cite{parameters} in understanding their relationships to underlying Coulombic interactions,
and in some systems also because of edge reconstructions\cite{edgereconstructions} or the role played by nearby metallic
gates\cite{gates}. In this paper we propose that measurement of edge magnetoplasmon\cite{edgemagnetoplasmons}  
properties in QHLJ systems can provide the required information experimentally.
To illustrate the potential of this approach, we derive analytic expressions for the 
edge magnetoplasmon spectrum of a simple QHLJ model and present numerical results for a more 
realistic model.  We find that as interactions across the barrier strengthen the independent 
magnetoplasmon excitations of the separated quantum Hall regions evolve into two quite 
distinct modes, a plasmon excitation localized along the barrier and a chiral magnetoplasmon 
mode that extends along the entire boundary of the compound system.  This change in 
character of the excitation spectrum, and accompanying changes in the strength of resonant absorption 
of EM radiation polarized either along or perpendicular to the barrier, can be used to reliably estimate
the strength and sign of interactions across the barrier.  

\noindent {\it Edge Magnetoplasmon Excitations of the QHLJ Model} --- 
We concentrate in this paper on the case of $\nu=1$ quantum Hall states for which 
the low-energy edge physics can, in the absence of reconstructions\cite{edgereconstructions}
be entirely\cite{wenedge} described in terms of the edge charge densities $\rho_L(x)$ and 
$\rho_R(x)$, of the left and right quantum Hall edges. 
The QHLJ model that we use to obtain a qualitative understanding of the 
magnetoplasmon spectrum includes interactions between left and 
right subsystems only in the barrier region:  
\begin{equation}
\begin{array}{l}
\displaystyle
H=H_0+H_i\\[1mm]
\displaystyle
H_0=\pi\hbar v_f\int_0^P dx\left[\rho_L(x)\rho_L(x)+\rho_R(x)
\rho_R(x)\right]\\[1mm]
\displaystyle
H_i=2\pi\hbar v_fg\int_0^P dx F(x)\rho_L(x)\rho_R(x)\\[1mm]
\end{array}
\label{h}
\end{equation}
where $H_0$ neglects interactions between separate edges, $v_f$ is the
chiral edge mode velocity of an isolated subsystem 
and $H_i$ describes interactions across the barrier.  The parameter 
$g$ is the ratio of interactions across the barrier to self-interactions on an
isolated edge and $P$ is the total perimeter of the individual quantum Hall regions.  We 
measure distances along the edges using the 
coordinate system explained in Fig.~\ref{fig:one} so that 
$F(x)\equiv \theta(x)-\theta(x-W)$ where $W$ is the width of the Hall bar.
For $\nu=1$ this model can be 
quantized by imposing the commutation relations\cite{wenedge}:
\begin{equation}
\begin{array}{l}
\displaystyle
[\rho_L(x),\rho_L(x')]=\left({i\over 2\pi}\right)\partial_x \delta(x-x')
\\[3mm]
\displaystyle
[\rho_R(x),\rho_R(x')]=-\left({i\over 2\pi}\right)\partial_x \delta(x-x')
\\[3mm]
\displaystyle
[\rho_L(x),\rho_R(x')]=0.
\end{array}
\end{equation}
It follows that the equations of motion for $\rho_L$ 
and $\rho_R$ are:
\begin{equation}
\begin{array}{l}
\displaystyle
-\partial_t\rho_L(x,t) = - v_f\partial_x[\rho_L(x,t)+gF(x)\rho_R(x,t)]\\[2mm]
\displaystyle
-\partial_t\rho_R(x,t) = v_f\partial_x[\rho_R(x,t)+gF(x)\rho_L(x,t)].
\end{array}
\label{eom}
\end{equation}

The right hand sides of Eqs.~\ref{eom} provide expressions for the current densities
on left and right edges, $j_L(x) = - v_f [\rho_L(x,t)+gF(x)\rho_R(x,t)]$ and 
$j_R(x) = v_f [\rho_R(x,t)+gF(x)\rho_L(x,t)]$. 
Eqs.~\ref{eom} are readily solved for fixed $g$ and solutions in the $g=0$ and 
$g \ne 0$ regions can be matched by requiring that the current densities be continuous\cite{safi,Agosta}. 
The general solution is most conveniently expressed in terms of fields
$\phi_{L,R}(x)$, related to the charge densities by $\rho_{L,R}
(x,t) = \pm \partial_x\phi_{L,R}(x,t)/(2 \pi)$.
The matching condition that $j_{L,R}$ should be continuous is equivalent to 
the condition that normal mode $\phi_{L,R}$ solutions at each frequency $\omega$ should be 
continuous.  The general form of a normal mode with frequency $\omega$ 
in the barrier region is
\begin{eqnarray} 
\phi_L(x) &=& A \exp(iqx) + B \exp(-iqx)  \nonumber \\
\phi_R(x) &=& A \; \frac{\sqrt{1-g^2}-1}{g}  \exp(iqx) \nonumber \\ 
&-&  B \; \frac{\sqrt{1-g^2}+1}{g} \exp(-iqx).
\label{phiofx} 
\end{eqnarray} 
where $q = \omega/(v_f \sqrt{1-g^2})$.  

The evolution of a unit
impulse charge traveling along a chiral edge can 
be understood on the basis of Eq.~\ref{phiofx}, charge conservation along both edges, the chirality of the outer edges, and 
the reduced wave velocity along the barrier. 
A unit impulse charge approaching the barrier from the left along the top of the 
Hall bar must launch an impulse that travels down along the barrier in order to 
conserve charge along the left edge.  Since this density wave travels at a slower 
velocity $v_I$, it must have left-edge charge density that is larger by the factor of 
$v_f/v_I = 1/\sqrt{1-g^2}$.
The equations of motion in the barrier region imply that the ratio of
right side charge to left side charge for a downward traveling barrier wave is 
$(\sqrt{1-g^2}-1)/g$ so that a downward traveling charge is also induced 
on the right hand side of the barrier.  To conserve right edge charge, a chiral
edge wave traveling at the higher velocity with charge $(1-\sqrt{1-g^2})/g$ 
must also be launched on the upper right.  This part of the edge density wave may be regarded as the  
chiral lead component that has been reflected by the barrier.  For a QHLJ system, however, 
is spatially separated from the incoming wave.
The total current traveling down the barrier
approaches $0$ for $g \to 1$ and $2 v_f$ for $g \to -1$.
These conclusion argued for here on the basis of charge conservation and equations of motion, also follow
from the matching conditions discussed previously.  Each of the waves launched 
by the original impulse will bifurcate when it intersects with an edge/barrier
boundary.
 
To determine the normal mode frequencies we integrate $\phi_L$ and $\phi_R$ around their
respective edges and apply the single-valuedness condition: 
\begin{equation}
\left(
\begin{array}{c}
\phi_L(P)\\
\phi_R(P)
\end{array}
\right)=
\left(
\begin{array}{cc}
U_{LL}&U_{LR}\\
U_{RL}&U_{RR}\\
\end{array}
\right)
\left(
\begin{array}{c}
\phi_L(0)\\
\phi_R(0)
\end{array}
\right)
=
\left(
\begin{array}{c}
\phi_L(0)\\
\phi_R(0)
\end{array} 
\right).
\label{propogator} 
\end{equation}
Normal modes occur when the determinant of $U_{I,J}-\delta_{I,J}$ vanishes.
In Eq.~\ref{propogator}, 
$U_{RR}^*=U_{LL}$, $U_{LR}^*=U_{RL}$, 
\begin{equation}
\begin{array}{c}
\displaystyle
U_{LL}=e^{iq_0P'}\left(\cos(q_IW)+ {i\over \sqrt{1-g^2}} \sin(q_IW)
\right) \nonumber \\[2mm]
\displaystyle
U_{LR}=i{g\over \sqrt{1-g^2}} e^{iq_0P'}\;\sin(q_IW),
\end{array}
\end{equation}
$q_0=\omega/v_f$ and $q_I=q_0 / \sqrt{1-g^2}$ are the local wavevectors in 
chiral and barrier regions, $P'=2L+W$ is the part of the perimeter of a quantum Hall region that 
is not along the barrier and $P'+W=P$ is the total perimeter.  We have, for 
illustrative purposes, assumed that 
the two quantum Hall regions are identical.  The
collective mode frequencies solve 
\begin{equation} 
\cos(q_0P') \cos(q_I W) - \frac{\sin(q_0P') \sin(q_IW)}{\sqrt{1-g^2}} =1. 
\label{freqeq}
\end{equation}  

\noindent {\it Weak Interaction Limit} --- In the absence of interactions across
the barrier ($q_I=q_0, g=0$) Eq.~\ref{freqeq} implies that the magnetoplasmon 
reference frequencies are given by $f_n = \omega_n /2\pi = n v_f/P$.  The period of 
the fundamental edge magnetoplasmon mode is just the transit time for an edge wave to 
move entirely around an individual incompressible region, implying that two independent modes
occur at each frequency.  Since typical edge magnetoplasmon velocities\cite{edgemagnetoplasmons} are 
$\sim 10^{6} {\rm m/s}$, frequencies are in the GHz range for samples with perimeters in the 
mm range.  Weak interactions shift and split the independent incompressible region 
eigenmodes.  From Eq. \ref{freqeq} we find that for small $g$
\begin{equation} 
f_n = v_f/P [n \pm g\sin(2\pi nL/P) / 2\pi + \ldots ] 
\label{smallg}
\end{equation} 
If the sample geometry is known $g$ can be extracted from a measurement of the mode splitting. 

\begin{figure}
\includegraphics{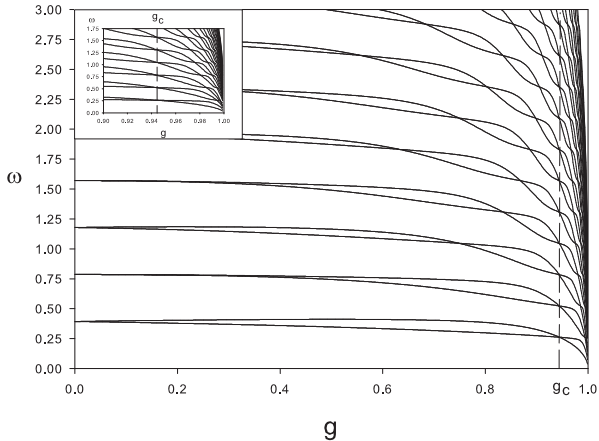}
\caption{\label{fig:two} The energy spectra calculated numerically are plotted
from $g=0$ to $1$. The spectrum is independent of the sign of $g$.
For this illustration we choose $W=L=4$, $v_f=1$, which implies that $g_c\approx 0.943$.
For $\vert g\vert<g_c$, mode splittings for each $n$ are 
clearly seen, while for $\vert g\vert>g_c$, the spectrum bends toward low 
energy and evolves into two sets of equally spaced modes with distinct fundamental
frequencies $f_0$ and $f_I$.  The crosses mark the commensurability positions where $f_0/f_I=m/n$ with $n,m$ 
both odd or both even.}
\end{figure}

\noindent {\it Strong Interaction Limit} --- For
$|g|$ can be close to $1$, the instability limit, the collective modes are determined by the conditions $\sin(q_0P')=0$ and 
$\sin(q_IW)=0$, leading to two sets of equally spaced modes with distinct fundamental frequencies at 
$f_0= v_f/2P'$ and $f_I=v_I/2W$. 
Apparently the edge magnetoplasmon modes in the strong 
interaction limit consist of independent modes localized along the barrier and along the 
chiral leads.  The periods $2P'/v_f$ and $2W/v_I$, respectively correspond to the transit
times for a wave traveling around the combined outer chiral edge at velocity 
$v_f$ and a wave traveling back and forth along the barrier at velocity $v_I$.  
Another interesting feature of the mode spectrum, illustrated in Fig. \ref{fig:two},
is the set of level crossings that occur at large values of $|g|$.  
The degeneracies occur when the ratio of $2P'/v_f$ to 
$2W/v_I$ is a rational number $n/m$ with $n$ and $m$ being both odd or both even 
integers, and occur at frequencies $\omega_l=(n \pi v_f/P')l$, $l=1,2,...$.

\noindent {\it Crossover Interaction Strength} --- 
$g^2=g_c^2=1-(W/P')^2$, the interaction strength for which $q_0P'=q_IW$
marks the crossover between weak and strong interaction limits. 
At $g=g_c$ the time required to travel at velocity $v_I$ along the barrier is 
equal to the time required to travel a distance $P'$ at velocity $v_f$ along one of the 
chiral edge loops.  In this case $f_0 = f_I$ and the two strong interaction modes 
become degenerate. For $g>g_c$, the lowest-energy
mode propagates primarily inside the interacting region, since the mode with 
lowest eigen frequency is the mode with longest period $T$. 

\noindent {\it Quantum Edge Theory} ---
The Hamiltonian in Eq. (1) can be solved quantum mechanically by bosonization.
We define two sets of bosonic fields:
\begin{equation}
\begin{array}{l}
\displaystyle
a_{L,k}\equiv \sqrt{{2\pi\over k P}}\rho_L(-k)\,\,\, , \,\,\,
a_{L,k}^\dagger\equiv \sqrt{{2\pi\over k P}}\rho_L(k)\\[3mm]
\displaystyle
a_{R,k}\equiv \sqrt{{2\pi\over k P}}\rho_R(k)\,\,\, , \,\,\,
a_{R,k}^\dagger\equiv \sqrt{{2\pi\over k P}}\rho_R(-k)\\[2mm]
\end{array}
\end{equation}
The Hamiltonian is readily expressed as a quadratic function of these bosonic fields.
The normal mode solutions to the semi-classical equation of 
motion map to the independent oscillators of the quantum model in the usual way.
The quantum version of the theory can be used to consider EM radiation absorption. 

\noindent
{\it Power Absorption} --- Edge states couple to the EM fields
via the Hamiltonian\cite{mahan}:
\begin{equation}
H_{ext}={i\over \omega}\int_0^P\,dx\,(\vec{j}_L+\vec{j}_R)\cdot \vec{E}
\end{equation}
where $\vec{j}_{L}, \vec{j}_{R}$ are current operators for left and right QH 
samples respectively and the EM waves of interest will typically have 
wavelengths much larger than the sample size. 
For simplicity we choose $W=L$ so that the geometry of the QH samples is 
square. The power absorption spectrum can be evaluated by Fermi's Golden Rule:
\begin{equation}
P(\omega)=2\pi\sum_{\alpha,\beta}\vert\langle\alpha\vert H_{ext}\vert\beta
\rangle\vert^2(E_\alpha-E_\beta)\delta(E_\alpha-E_\beta-\hbar\omega).
\end{equation}
Peaks in the power absorption spectrum correspond to resonantly 
excited EMPs.  

\begin{figure}
\includegraphics{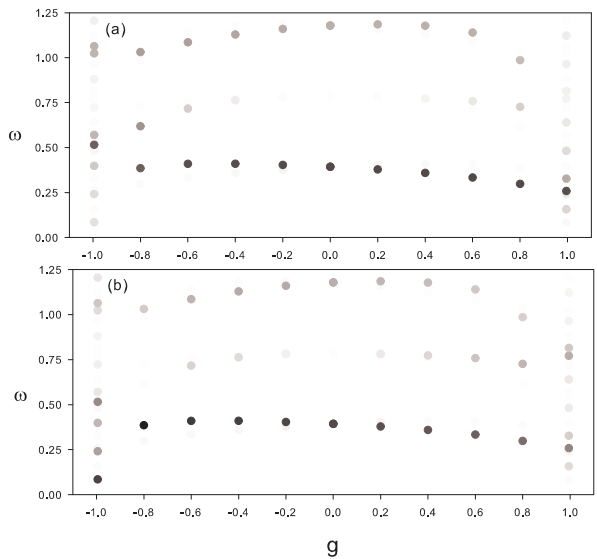}
\caption{\label{fig:three} Resonant mode oscillator strength 
for $g=0$, $g=\pm 0.2$, $g=\pm 0.4$, $g=\pm 0.6$, $g=\pm 0.8$, and
$g=\pm 0.995$, with EM fields polarized (a) perpendicular to (b) 
along the narrow barrier. The strength of the each resonant 
absorption peak is indicated by the darkness of the circle.}
\end{figure}

Numerical results for the absorption spectrum are shown in Fig. \ref{fig:three}. In general only the lowest few modes tend to have substantial absorption 
strength, because higher energy modes correspond to current oscillations with 
larger spatial varation leading to smaller matrix elements after integration along 
the edges. Although the energy spectrum does not 
depend on the sign of $g$, the mode absorption strengths at $g$ and $-g$ does. 
For weak interactions the sharpest peak is for the lowest energy mode for repulsive 
interaction and at the {\it second} lowest energy mode for attractive 
interaction. This feature can be understood from the properties of the normal modes. 
For repulsive interactions the 
lowest-energy EMP has opposite charge densities on left and right edges.
Because currents flow in opposite directions in left and right edges,
the net current is nonzero and coupling to the EM field does not vanish.
For attractive interaction the energetic ordering of the two levels switches. 
The strongest absorption then occurs for the second lowest energy EMP. 

The polarization of EM fields also influences the strength of peaks. It can be 
seen in Fig. \ref{fig:three} that for repulsive (attractive) interactions, 
the peak with EM fields orienated along the barrier line is weaker 
(stronger) than for polarization perpendicular to the barrier line. This
occurs because the current along the barrier is enhanced for 
attractive interactions and suppressed for repulsive interactions.

In the strong interaction limit, the localized EMPs modes propagate mainly in the
barrier region and are not probed by EM fields polarized perpendicular to the 
barrier. As argued in previous sections, the current of those localized EMPs 
is small for repulsive interactions but not for attractive interactions.
This property explains why EM fields
applied along the barrier, produce strong absorption at the lowest 
energy modes for $g=-0.995$ but not for $g=0.995$ in Fig. \ref{fig:three}(b).

\noindent
{\it Comparison of Coulomb model and Toy model} --- A more realistic model can
be obtained by replacing $H_i$ in Eqs. \ref{h} with both intra- and inter-edge
Coulomb interactions.  Since interactions on single isolated incompressible 
system edge to a good approximation influence only the mode velocity\cite{parameters}
and the inter-edge Coulomb interaction is much stronger in the barrier 
region than on other portions of the edge,  we can expect the 
simple model properties to be similar to the realistic case.  We have solved the 
more realistic Coulomb model numerically with the  
relative strength of inter-edge Coulomb interaction interactions which can be 
tuned by changing the separation distance between left and right edges to verify this expectation.
We conclude that the simple QHLJ model should be very useful for the 
interpretation of experimental results.  

\noindent
{\it Acknowledgements} --- This work was supported by the Welch Foundation.  The authors
acknowledge helpful interactions with Matt Grayson, Woowon Kang, and Vittorio Pellegrini.

\end{document}